\documentclass[twocolumn]{aastex61}

\usepackage{graphicx}
\usepackage{subfigure}

\newcommand{\bd}{\begin{displaymath}}
\newcommand{\ed}{\end{displaymath}}
\newcommand{\be}{\begin{equation}}
\newcommand{\ee}{\end{equation}}
\newcommand{\beaa}{\begin{eqnarray*}}
\newcommand{\eeaa}{\end{eqnarray*}}
\newcommand{\bea}{\begin{eqnarray}}
\newcommand{\eea}{\end{eqnarray}}

\received{\today}
\revised{\today}
\accepted{\today}
\submitjournal{ApJ}

\shorttitle{Measuring the value of the Hubble constant ``\`a la Refsdal''}
\shortauthors{Grillo et al.}

\begin{document}

\title{Measuring the value of the Hubble constant ``\`a la Refsdal''}

\correspondingauthor{Claudio Grillo}
\email{claudio.grillo@unimi.it}

\author[0000-0002-5926-7143]{C.~Grillo}
\affiliation{Dipartimento di Fisica, Universit\`a  degli Studi di Milano, via Celoria 16, I-20133 Milano, Italy}
\affiliation{Dark Cosmology Centre, Niels Bohr Institute, University of Copenhagen, Juliane Maries Vej 30, DK-2100 Copenhagen, Denmark}


\author{P.~Rosati}
\affiliation{Dipartimento di Fisica e Scienze della Terra, Universit\`a degli Studi di Ferrara, Via Saragat 1, I-44122 Ferrara, Italy}
\affiliation{INAF - Osservatorio Astronomico di Bologna, via Gobetti 93/3, I-40129 Bologna, Italy}

\author{S.~H.~Suyu}
\affiliation{Max-Planck-Institut f{\"u}r Astrophysik, Karl-Schwarzschild-Str. 1, 85748 Garching, Germany}
\affiliation{Institute of Astronomy and Astrophysics, Academia Sinica, P.O. Box 23-141, Taipei 10617, Taiwan}
\affiliation{Physik-Department, Technische Universit{\"a}t M{\"u}nchen, James-Franck-Straße 1, 85748 Garching, Germany}

\author{I.~Balestra}
\affiliation{University Observatory Munich, Scheinerstrasse 1, D-81679 Munich, Germany}

\author{G.~B.~Caminha}
\affiliation{Kapteyn Astronomical Institute, University of Groningen, Postbus 800, 9700 AV Groningen, the Netherlands}

\author{A.~Halkola}
\affiliation{ }

\author{P.~L.~Kelly}
\affiliation{Department of Astronomy, University of California, Berkeley, CA 94720-3411, USA}

\author{M.~Lombardi}
\affiliation{Dipartimento di Fisica, Universit\`a  degli Studi di Milano, via Celoria 16, I-20133 Milano, Italy}

\author{A.~Mercurio}
\affiliation{INAF - Osservatorio Astronomico di Capodimonte, Via Moiariello 16, I-80131 Napoli, Italy}

\author{S.~A.~Rodney}
\affiliation{Department of Physics and Astronomy, University of South Carolina, 712 Main St., Columbia, SC 29208, USA}

\author{T.~Treu}
\affiliation{Department of Physics and Astronomy, University of California, Los Angeles, CA 90095, USA}

\begin{abstract}
  Realizing Refsdal's original idea from 1964, we present estimates of the Hubble constant that are complementary to and potentially competitive with those of other cosmological probes. We use the observed positions of 89 multiple images, with extensive spectroscopic information, from 28 background sources and the measured time delays between the images S1-S4 and SX of supernova ``Refsdal'' ($z = 1.489$), which were obtained thanks to \emph{Hubble Space Telescope} (\emph{HST}) deep imaging and Multi Unit Spectroscopic Explorer (MUSE) data. We extend the strong lensing modeling of the Hubble Frontier Fields (HFF) galaxy cluster MACS J1149.5$+$2223 ($z = 0.542$), published by \citet{gri16}, and explore different $\Lambda$CDM models. Taking advantage of the lensing information associated to the presence of very close pairs of multiple images at various redshifts and to the extended surface brightness distribution of the SN Refsdal host, we can reconstruct the total mass density profile of the cluster very precisely. The combined dependence of the multiple image positions and time delays on the cosmological parameters allows us to infer the values of $H_{0}$ and $\Omega_{\rm m}$ with relative (1$\sigma$) statistical errors of, respectively, 6\% (7\%) and 31\% (26\%) in flat (general) cosmological models, assuming a conservative 3\% uncertainty on the final time delay of image SX and, remarkably, no priors from other cosmological experiments. Our best estimate of $H_{0}$, based on the model described in this work, will be presented when the final time-delay measurement becomes available. Our results show that it is possible to utilize time delays in lens galaxy clusters as an important alternative tool for measuring the expansion rate and the geometry of the Universe.

\end{abstract}

\keywords{gravitational lensing: strong --- cosmological parameters --- distance scale --- galaxies: clusters: individuals: MACS J1149.5$+$2223 --- dark matter --- dark energy}

\section{Introduction}
\label{sec:intro}

The Hubble constant ($H_{0}$) is a fundamental cosmological parameter that defines many of the most important scales in the Universe: its size, age, expansion rate, and critical density. In the past 25 years, remarkable progress has been made on the determination of the value of $H_{0}$, thanks to the observations of 
Cepheids and supernovae (SNe; e.g., \citealt{FreedmanEtal01}; \citealt{fre12}; \citealt{rie16}), the cosmic microwave background (e.g., \citealt{hin13}; \citealt{pla16}), water masers (e.g., \citealt{rei13}; \citealt{kuo15}; \citealt{gao16}), and quasars (QSOs) strongly lensed by galaxies (e.g., \citealt{suy13,suy14}; \citealt{won17}). More recently, the combination of gravitational wave and electromagnetic data has proved to be a promising new way to estimate the Hubble constant (\citealt{abb17}; \citealt{gui17}).

The increased precision
of the most recent measurements of $H_{0}$ based on the distance ladder ($73.24 \pm 1.74$ km s$^{-1}$ Mpc$^{-1}$; \citealt{rie16}) and from the \emph{Planck} satellite ($67.74 \pm 0.46$ km s$^{-1}$ Mpc$^{-1}$; \citealt{pla16}) has revealed some tension at the $\approx$3$\sigma$ level (see also \citealt{rie18}). This might point to the presence of unknown systematic effects or interesting new physics. To clarify this situation, the results of additional independent and high-precision techniques, which rely on different physics, are fundamental.

As theoretically predicted by \citet{ref64}, strongly lensed SNe with measured time delays between the multiple SN images provide an independent way to measure the Hubble constant.  Given the rarity of lensed SNe, the strong lens time delay method has been utilised with lensed quasars until now.  In particular, the H0LiCOW program (\citealt{suy17}), together with the COSMOGRAIL program (e.g., \citealt{TewesEtal13a}; \citealt{CourbinEtal17}), aims to measure $H_0$ with $<3.5\%$ uncertainty from the joint analysis of five different lensing systems (see \citealt{bon17} for the initial results from three lenses; $H_{0} = 71.9^{+2.4}_{-3.0}$ km s$^{-1}$ Mpc$^{-1}$).  SN ``Refsdal'' ($z = 1.489$) was discovered by \citet{kel15,kel16a} to be strongly lensed by the Hubble Frontier Fields (HFF) galaxy cluster MACS J1149.5$+$2223 (hereafter MACS 1149; $z = 0.542$; \citealt{tre15b}; \citealt{gri16}, hereafter G16). We show here that by using a full strong lensing analysis of the first multiply-imaged and spatially-resolved SN Refsdal, which includes its time-delay measurements (\citealt{kel16}; \citealt{rod16}) and a robust knowledge of the cluster gravitational potential derived from a large number of multiple images, it is possible to measure the values of the Hubble constant and of cosmological parameters with a precision comparable to that of other standard techniques.

\section{Methods}
\label{sec:met}

\subsection{Theory}
\label{sec:theo}

In this section, we introduce very concisely the dependence of some of the observables related to the multiple images of a lensed source on the values of the cosmological parameters. For more details about the general theory of gravitational lensing, we refer to dedicated textbooks (e.g., \citealt{sch92}; \citealt{dod17}).

If a source is strongly lensed into two images, $\rm i_{1}$ and $\rm i_{2}$, the difference in time that light takes to reach the observer from the two different directions, i.e. the time delay between the two images, $\Delta t_{\rm i_{1}i_{2}}$, is 
\begin{equation}
\label{eq:2}
\Delta t_{\rm i_{1}i_{2}} = \frac{D_{\Delta t}}{c} \Delta \phi_{\rm i_{1}i_{2}},
\end{equation}
where $\phi$ is the Fermat potential (connected to the gravitational potential of the lens; see \citealt{sch92}) and $D_{\Delta t}$ is the time-delay distance (see \citealt{suy10b}), defined as
\begin{equation}
\label{eq:1}
D_{\Delta t}(z_{\rm d},z_{\rm s}) = (1+z_{\rm d})\frac{D_{\rm d}D_{\rm s}}{D_{\rm ds}},
\end{equation}
with $z_{\rm d}$ and $z_{\rm s}$ as the redshifts of the deflector and the source, respectively, and $D_{\rm d}$, $D_{\rm s}$, and $D_{\rm ds}$ as the observer-deflector, observer-source and deflector-source angular-diameter distances, respectively. The ratio of the three angular-diameter distances entering in Equation (\ref{eq:1}) implies that $D_{\Delta t} \propto H_{0}^{-1}$. From Equations (\ref{eq:2}) and (\ref{eq:1}), it follows that if the time delay between two images of the same source can be measured observationally and the Fermat potential reconstructed through strong lensing modeling, then the value of the time-delay distance, thus those of the cosmological parameters, can be constrained. In general, $D_{\Delta t}$ can be determined more precisely if the time delays between more than two multiple images are available.

If a lens produces multiple images of two sources, located at different redshifts $z_{\rm s_{1}}$ and $z_{\rm s_{2}}$, the observed positions of the multiple images provide information about the total mass profile of the lens and the so-called family ratio (e.g., \citealt{sou04}; \citealt{jul10}):
\begin{equation}
\label{eq:3}
\Xi(z_{\rm d},z_{\rm s_{1}},z_{\rm s_{2}}) = \frac{D_{\rm ds_{1}}D_{\rm s_{2}}}{D_{\rm s_{1}}D_{\rm ds_{2}}}.
\end{equation}
Depending on the complexity of the lens mass model and on the number of observed multiple images, Equation (\ref{eq:3}) shows that a ratio of ratios of angular-diameter distances can, in principle, be estimated, and from that the values of the relevant cosmological parameters inferred. This method can be employed effectively in lens galaxy clusters with a large number of spectroscopically confirmed multiple images, where different values of $\Xi$ can be used at the same time, as recently illustrated by \citet{cam16} (see \citealt{joh16} and \citealt{ace17} for further discussion).

We note that (1) time-delay distances are primarily sensitive to the value of $H_{0}$, and more mildly on other cosmological parameters (see also \citealt{lin11}); (2) a lensing system with several multiply imaged sources can provide constraints on the value of $\Omega_{m}$ and $\Omega_{\Lambda}$ 
(in $\Lambda$CDM), 
but it is insensitive to the value of $H_{0}$ (the value of the Hubble constant cancels out in the ratio of Equation (\ref{eq:3})). The ideal cosmological laboratory is then a lens with a relatively simple total mass distribution accurately constrained by many bona fide multiple images from sources covering a wide redshift range, some of which are time-varying (i.e., allowing time delay measurements). MACS 1149 with SN 
Refsdal 
provides such a laboratory (e.g., \citealt{smi09}; \citealt{zit09}).

In the following strong-lensing analysis, the total chi-square $\chi^{2}_{\rm tot}$ (or, equivalently, the likelihood) results from the sum of two different terms: $\chi^{2}_{\rm pos}$ and $\chi^{2}_{\rm td}$. The former and the latter quantify the agreement between the observed and model-predicted values of the multiple-image positions and time delays, respectively, weighted by the corresponding observational uncertainties. We note that the model-predicted values of the time delays in $\chi^{2}_{\rm td}$ are calculated at the model-predicted positions of the multiple images (this is more appropriate in lens galaxy clusters, where the observed multiple image positions differ on average from the model-predicted ones by $\approx$0.5\arcsec).

\begin{figure*}
\centering
\includegraphics[width=0.98\textwidth]{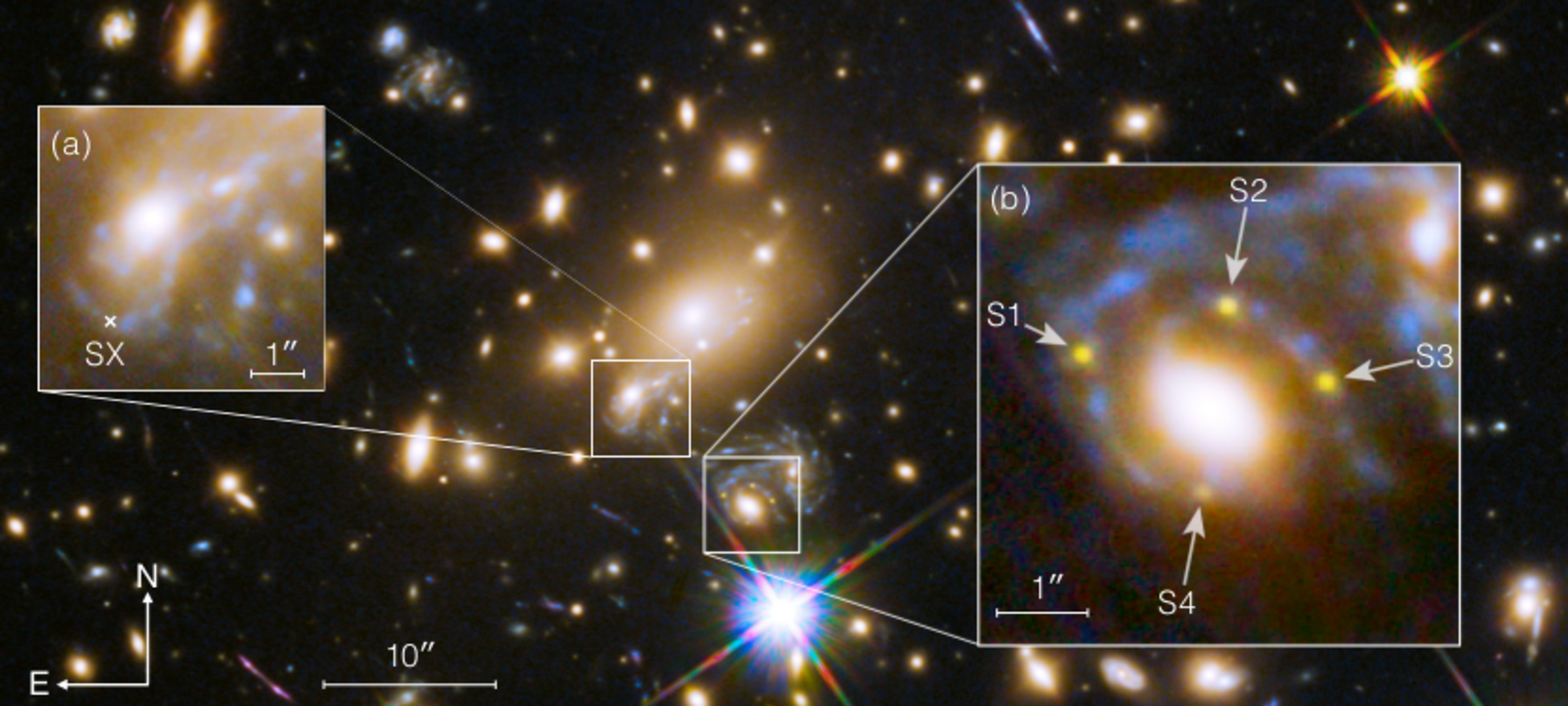}
\caption{Positions of the five detected images of SN Refsdal. The background image shows the MACS 1149 cluster, combining imaging from the Hubble Space Telescope in optical and near-infrared bands with effective wavelengths spanning 4350 \AA $\,$ to 16000 \AA.  Inset panel (a) shows the location of the fifth detected image, SX, first observed in December 2015.  Inset panel (b) shows the first four images, S1--S4, which were apparent when SN Refsdal was first detected in November 2014. (Original image credit: NASA, ESA/Hubble)}
\label{fig3}
\end{figure*}

\subsection{Lens modeling}
\label{sec:mod}
We summarize here the details of the strong-lensing modeling of MACS 1149 presented in G16. We 
extend the absolute best-fitting cluster mass model (labelled as MLV G12F) to include the time delays of the multiple images of SN 
Refsdal 
and to let the values of the cosmological parameters free to vary. The interested reader is referred to 
G16
for a more extensive description of the modeling and statistical analysis and to a similar work on another HFF cluster, MACS J0416.1$-$2403 (\citealt{gri15}, hereafter G15). The software used to model these clusters is {\sc Glee}, developed by A.~Halkola and S.~H.~Suyu \citep{SuyuHalkola10,SuyuEtal12}. {\sc Glee} has already been employed to study the mass distribution of lens galaxies and galaxy clusters and to probe the expansion history of the Universe through lensed quasars (e.g., \citealt{suy13,suy14}; \citealt{won17}). 

\begin{table*}
\centering
\caption{Strong-lensing models with the corresponding adopted time delays for the multiple images of SN 
Refsdal 
and their values of the best-fitting $\chi^2$ (mininum-$\chi^2$), for the multiple image positions ($\chi^{2}_{\rm pos}$), time delays ($\chi^{2}_{\rm td}$) and total ($\chi^{2}_{\rm tot}$), and degrees of freedom (dof). Flat $\Lambda$CDM models ($\Omega_{m}+\Omega_{\Lambda} = 1$) with uniform priors on the values of the 
 cosmological parameters ($H_{0} \in [20,120]$ km~s$^{-1}$~Mpc$^{-1}$ and $\Omega_{m} \in [0,1]$) are considered.}
\begin{tabular}{ccccccccc}
\hline\hline \noalign{\smallskip}
ID & $\Delta t_{\rm S2:S1}$$^{\mathrm{a}}$ & $\Delta t_{\rm S3:S1}$$^{\mathrm{a}}$ & $\Delta t_{\rm S4:S1}$$^{\mathrm{a}}$ & $\Delta t_{\rm SX:S1}$$^{\mathrm{b}}$ & $\chi^{2}_{\rm pos}$ & $\chi^{2}_{\rm td}$ & $\chi^{2}_{\rm tot}$ & dof \\
 & (days) & (days) & (days) & (days) & & & & \\
\noalign{\smallskip} \hline \noalign{\smallskip}
${\rm \Delta t(t)}$ & $4 \pm 4$ & $2 \pm 5$ & $24 \pm 7$ & $345 \pm 10$ & 88.1 & 1.4 & 89.5 & 93 \\
${\rm \Delta t(p)}$ & $7 \pm 2$ & $0.6 \pm 3$ & $27 \pm 8$ & $345 \pm 10$ & 88.9 & 1.2 & 90.1 & 93 \\
\noalign{\smallskip} \hline
\end{tabular}
\begin{list}{}{}
\item[$^{\mathrm{a}}$]Measured by \citet{rod16}.
\item[$^{\mathrm{b}}$] Preliminary estimate with a conservative uncertainty (based on \citealt{kel16}).
\end{list}  
\label{models}
\end{table*}

\subsubsection{Lensing observables}
\label{sec:obs}

We optimize the strong-lensing model (cluster mass and cosmological) parameters 
with uniform priors,
over the positions of 89 observed and reliable multiple images belonging to 10 different sources ($1.240 \le z \le 3.703$) and to 18 knots of the SN 
Refsdal
host ($z = 1.489$), further validated by MUSE velocities from [\ion{O}{2}] emission (see Figures 8 and 9 in G16 and \citealt{dit18}). As detailed in G16, the considered positional uncertainty of each image is 0.26\arcsec $\,$ in order to get a $\chi^{2}$ value that is comparable to the number of the degrees of freedom (except for the 5 multiple images of SN Refsdal, S1-S4 and SX, for which we use 0.13\arcsec). The redshift values of the 7 spectroscopically confirmed multiply-imaged sources are fixed, while the remaining 3 systems are included with a uniform prior on their redshifts, $z \in [0,6]$ (see G16). Moreover, we include the observed time delays (and their statistical errors) of the images S2, S3, and S4, relative to S1, of SN Refsdal, as measured from their full light curves by using a set of templates (t) or polynomials (p) (see \citealt{rod16}). We select a fiducial time delay for SX of 345 days based on the broad constraints presented in Figure~3 of \citet{kel16}, where only the first photometric points of the light curve of this last image were used. For this quantity, we first use a conservative statistical uncertainty of 10 days, corresponding to approximately a $3\%$ error, and then consider 7 and 4 days (the final observational error is expected to be $\approx$1-2\%, thus closer to these last cases; see \citealt{rod16}; \citealt{kel16}). To accommodate possible differences in the ultimate measurement of the SX time delay, we also test the effect of conservative positive and negative shifts of 15 and 30 days, i.e. 375, 360, 330, and 315 days, with a fixed uncertainty of 10 days.

\subsubsection{Cluster mass components}
\label{sec:comp}

The absolute best-fitting cluster mass model (MLV G12F) presented in G16 contains three extended dark-matter halo components, modeled as cored elliptical pseudo-isothermal mass distributions, and a highly pure sample of 300 candidate cluster members, in the form of dual pseudo-isothermal mass distributions. Of the 300 candidate cluster members (55\% spectroscopically confirmed), 298 are approximated as axially symmetric and with vanishing core radius and scaled with total mass-to-light ratios increasing with their near-IR (\emph{HST} F160W) luminosities (as suggested by the tilt of the Fundamental Plane; \citealt{fab87,ben92}), and 2 are elliptical with mass parameters free to vary (the closest galaxies, in projection, to the SN Refsdal multiple images; see G16). The results of this particular model are also the ones used in the comparative study by \citet{tre15b}.

\subsubsection{Cosmological models}
\label{sec:cosmo}

We consider flat ($\Omega_{m}+\Omega_{\Lambda} = 1$; see Table \ref{models} and Figures \ref{fig1} and \ref{fig2}) and general (see Figure \ref{fig2}) $\Lambda$CDM models with uniform priors on the values of the considered cosmological parameters: $H_{0} \in [20,120]$ km~s$^{-1}$ Mpc$^{-1}$ and either $\Omega_{m} \in [0,1]$ or $\Omega_{m} \in [0,1]$ and $\Omega_{\Lambda} \in [0,1]$, respectively.

\section{Results}
\label{sec:res}

We sample the posterior probability distribution function of the parameters of the lensing models using a standard Bayesian analysis and \texttt{emcee} (\citealt{for13}; for more general details, see also Sect.~3.2 in G15 and Sect.~4.4 in G16). We get $\chi^{2}$ values (see Table \ref{models}) that are comparable to the number (93) of the degrees of freedom (dof). The latter is given by the difference between the number of lensing observables (178 $x$ and $y$ coordinates of the multiple images and 4 time delays for S2, S3, S4, and SX) and that of the model free parameters (28 describing the cluster total mass distribution, 56 $x$ and $y$ coordinates of lensed sources, 3 redshifts of the photometric families and 2 for $H_{0}$ and $\Omega_{m}$ 
in flat $\Lambda$CDM). In this way, possible small dark-matter substructures, deviations from elliptical mass profiles and some scatter in the adopted scaling relations for the cluster members, which have not been explicitly included in our model, are statistically taken into account, and realistic errors on the values of the model parameters can be estimated. We obtain final MCMC chains with approximately $8 \times 10^{5}$ samples for each model.

We have checked the values of the best-fitting ${\rm \Delta t(t)}$ model against those of the MLV G12F model in G16. The values associated to the cluster total mass distribution show differences that are on average on the order of 0.5$\sigma$, thus the two models are statistically consistent. This is not very surprising, since our blind predictions of the position, flux and time delay of SX, published in G16, were obtained there in a flat $\Lambda$CDM model ($\Omega_{m}$ = 0.3) with $H_{0}$ = 70 km s$^{-1}$ Mpc$^{-1}$ and were shown to be in very good agreement with the following observations (see \citealt{kel16}). The observables included in the ${\rm \Delta t(t)}$ model differ from those of the MLV G12F model essentially only in the inclusion of the measured position and time delay estimate of the multiple image SX by \citet{kel16}. The best-fitting cosmological values of the ${\rm \Delta t(t)}$ model are $H_{0}$ = 70.4 km s$^{-1}$ Mpc$^{-1}$ and $\Omega_{m}$ = 0.31. As a consequence, the cluster total mass distribution is not significantly different. We remark that this is not a circular argument, but only the demonstration that all the results are consistent.

\begin{figure}
\centering
\includegraphics[width=0.485\textwidth]{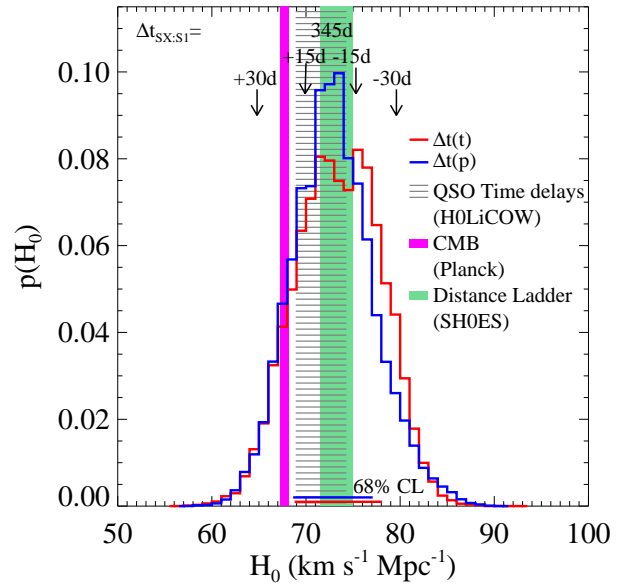}
\caption{Marginalized probability distribution functions of $H_{0}$. The results of the flat $\Lambda$CDM models listed in Table~\ref{models} are shown by the red and blue histograms (on the bottom, the corresponding 68\% CL intervals). The vertical arrows show the inferred median values of $H_{0}$ (in the (p) case) if the time delay of SX differs from 345 days by 15 or 30 days. Credible intervals, at 1$\sigma$ CL, from H0LiCOW (\citealt{bon17}), Planck (\citealt{pla16}) and SH0ES (\citealt{rie16}) are indicated in gray, magenta and green, respectively.}
\label{fig1}
\end{figure}

We show in Figure \ref{fig1} and Table \ref{stat} the posterior probability distribution function and the 1$\sigma$, 2$\sigma$ and 3$\sigma$ credible intervals of $H_{0}$, marginalized over all the other strong-lensing model parameters. We notice that the results obtained with the time delays of the images S2, S3, and S4 measured with a set of light curve templates (t) or with polynomials (p) are consistent, given the statistical uncertainties. Remarkably, we can infer the value of $H_{0}$ with a (1$\sigma$) statistical error of approximately $6\%$. If the statistical uncertainty on the SX time delay is 2\% (7 days) or 1\% (4 days), in both cases the statistical error on $H_{0}$ reduces only slightly to approximately $5\%$. If the true time delay for SX is longer (shorter) by $\approx4$\% or 9\% (i.e., 15 or 30 days), then this will translate into a value of $H_{0}$ which is smaller (larger) by approximately the same percentage. A simple linear interpolation of these values (in the (p) case) provides the following scaling result for $H_{0}$: $[72.5 - 0.233\textrm{d}^{-1} \times (\Delta t_{\rm SX:S1} - 345\textrm{d})]$ km s$^{-1}$ Mpc$^{-1}$. Interestingly, the value of $\Omega_{m}$ is on average inferred with a (1$\sigma$) statistical error of $\approx$30\% (the median value is not significantly affected by the precise value of the SX time delay), and is in excellent agreement with measurements based on geometrical and/or structure growth rate methods (see \citealt{pla16}).

\begin{figure}
\centering
\includegraphics[width=0.485\textwidth]{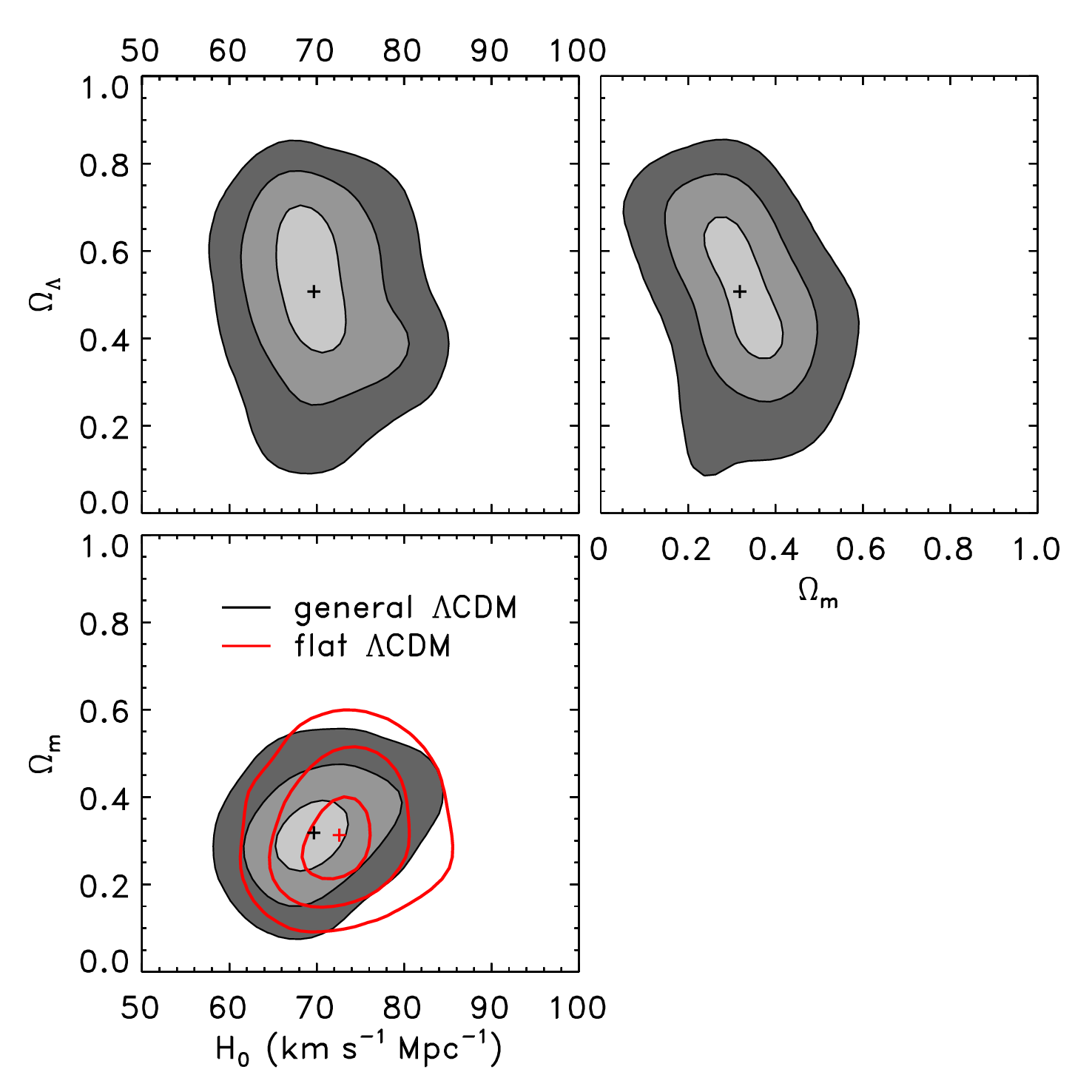}
\caption{Constraints on the cosmological parameters. The model ${\rm \Delta t(p)}$ (see Table \ref{models}) in flat (red) and general (gray) $\Lambda$CDM models with uniform priors on the values of the cosmological parameters ($H_{0} \in [20,120]$ km~s$^{-1}$ Mpc$^{-1}$, $\Omega_{m} \in [0,1]$ and $\Omega_{\Lambda} \in [0,1]$) is shown here. The cross symbols and the contour levels on the planes represent, respectively, the median values and the 1$\sigma$, 2$\sigma$, and 3$\sigma$ confidence regions, as obtained from MCMC analyses.}
\label{fig2}
\end{figure}

In Figure \ref{fig2}, we illustrate the inference on the values of the cosmological parameters $H_{0}$, $\Omega_{m}$ and $\Omega_{\Lambda}$. If we consider the second model (${\rm \Delta t(p)}$) of Table \ref{models} in a general $\Lambda$CDM model and vary all the strong-lensing model parameters at the same time, we obtain the following notable 1$\sigma$ confidence level (CL) constraints: $69.8^{+5.3}_{-4.1}$ km s$^{-1}$ Mpc$^{-1}$ for $H_{0}$, $0.32^{+0.08}_{-0.08}$ for $\Omega_{m}$ and $0.51^{+0.16}_{-0.15}$ for $\Omega_{\Lambda}$. This corresponds to relative statistical errors of approximately $7\%$, $26\%$ and $31\%$. The high precision on the values of $\Omega_{m}$ and $\Omega_{\Lambda}$ can be ascribed to the combination of constraints coming from the time delays and the multiple image positions of sources at different redshifts (see Equations (\ref{eq:1}) and (\ref{eq:3})).

\begin{table}
\centering
\caption{Median values and intervals at 1$\sigma$, 2$\sigma$, 3$\sigma$ confidence level of the Hubble constant $H_{0}$ (in km s$^{-1}$ Mpc$^{-1}$) for the models shown in Table \ref{models}.}
\begin{tabular}{ccccc}
\hline\hline \noalign{\smallskip}
ID & $H_{0}$ & 1$\sigma$ & 2$\sigma$ & 3$\sigma$ \\
\noalign{\smallskip} \hline \noalign{\smallskip}
${\rm \Delta t(t)}$ & 73.5 & $^{+4.6}_{-4.7}$ & $^{+8.4}_{-8.8}$ & $^{+12.4}_{-13.1}$ \\
${\rm \Delta t(p)}$ & 72.8 & $^{+4.3}_{-4.1}$ & $^{+9.5}_{-8.0}$ & $^{+14.1}_{-11.5}$ \\
\noalign{\smallskip} \hline
\end{tabular}
\label{stat}
\end{table}

\section{Discussion}
\label{sec:disc}


The term associated to the time delays ($\chi^{2}_{\rm td}$) gives a relatively small contribution (see Table \ref{models}) to the total chi-square value ($\chi^{2}_{\rm tot}$) of the best-fitting models, but we have checked that the former increases extremely rapidly if we vary the value of $H_{0}$. This explains why a precise estimate of the value of the Hubble constant is possible through this method. It is clear that within our modeling assumptions, the time-delay measurement of the latest image (SX) of SN Refsdal 
drives the inferred value and the error budget on $H_{0}$. In Figure \ref{fig1}, one can see that estimates of $H_{0}$ from strong-lensing analyses in galaxy clusters containing a large fraction of spectroscopically confirmed multiple images and one time-variable system, like SN 
Refsdal, 
could represent a noteworthy independent measurements to those obtained from lensed quasars ($H_{0}=71.9^{+2.4}_{-3.0}$ km s$^{-1}$ Mpc$^{-1}$; H0LiCOW, \citealt{bon17}), SNe distance ladder ($H_{0}=73.24 \pm 1.74$ km s$^{-1}$ Mpc$^{-1}$; SH0ES, \citealt{rie16}) and CMB ($H_{0}=67.74 \pm 0.46$ km s$^{-1}$ Mpc$^{-1}$; \citealt{pla16}; see also \citealt{hin13}) data.

Time-delay distances obtained from quasars multiply lensed by galaxies
have already provided very precise estimates of the value of $H_{0}$
and, when combined with independent probes, can also constrain other
cosmologically relevant quantities (\citealt{suy13,suy14};
\citealt{won17}; \citealt{bon17}). Several studies (e.g., \citealt{bir16}; \citealt{tre16}; \citealt{suy17}) have recognized that, in addition to
the spectroscopic redshifts of the lens and the source, the most
important steps toward accurate and precise cosmological measurements
through $D_{\Delta t}$ inference in Equation (\ref{eq:1}) are: (1) precise time delays, (2) high-resolution
images of the lensed sources, (3) precise stellar kinematics of the
lens galaxy, and (4) detailed information about the lens
environment. 
Long-term
(several years) monitoring campaigns of lensed quasars with either optical telescopes, notably by the COSMOGRAIL collaboration \citep[e.g.,][]{TewesEtal13a, CourbinEtal17}, or radio observations \citep[e.g.,][]{FassnachtEtal02}, together with advances in light-curve analysis techniques (e.g., \citealt{TewesEtal13b}; \citealt{hoj13}), have yielded precise time delays.
To convert these delays to $D_{\Delta t}$, an accurate lens mass model is needed, particularly concerning the radial mass density profile.
Steeper profiles yield larger Fermat potential
differences between two images, resulting in shorter estimated $D_{\Delta t}$, and thus larger inferred values of $H_{0}$
(\citealt{wuc02}; \citealt{koc02}). Moreover, in addition to the main
lens, there could be external mass contributions, associated to other
galaxies belonging to the same group of the main lens or to structures
along the line of sight. If not properly taken into account, this term
represents another important source of systematic error, the so-called
``mass-sheet degeneracy'' (\citealt{fal85}; \citealt{sch13}), in the
model prediction of the time delays. This explains why the surface
brightness reconstruction of multiple images, the use of independent
mass probes (e.g., through dynamical modeling; see \citealt{tre02}) for the main lens, and
a full characterization of its environment (i.e., points (2), (3), and
(4) mentioned above) are so relevant to a very accurate lens mass
model, thus to the success of this cosmological tool \citep[e.g.,][]{suy14, bir16, McCullyEtal17, RusuEtal17, ShajibEtal17, SluseEtal17, TihhonovaEtal17}.

In contrast to quasars, the time variability curve of a SN is much
simpler to model. For SN Refsdal, dedicated \emph{HST} monitoring
programs have already measured the time delays of the multiple images
S2-S4, relative to S1, and are expected to deliver soon a relative
precision of $\approx$1-2\% on the time delay of SX (\citealt{rod16};
\citealt{kel16}; \emph{HST} GO-14199). Furthermore, despite being more complex than that of
an isolated galaxy, the strong lensing modeling of MACS 1149 presents
some advantages. First, the identification of several multiply-lensed
knots in the SN Refsdal host (see Table 3 and Figure 7 in G16),
some of which are very close to the brightest cluster galaxy and radially
elongated, provides important information about the slope of the total
mass density profile of the cluster (see e.g., \citealt{cam17}). Then, the presence of several
pairs of angularly close multiple images (e.g., systems 2, 5, 6, 8,
and 14 in Table 2 of G16), from sources at different
redshifts, constrains tightly the lens tangential critical curves,
thus offering precise calibrations of the projected total mass of the
cluster within different apertures. In MACS 1149, these two rare coincidences reduce the need to include in the modeling information from a different total mass diagnostic, such as stellar dynamics in lens galaxies. In addition, the large number of
secure and spectroscopically confirmed multiple images observed in
galaxy clusters allows one to test different mass models and to choose
the best one (i.e., the best parametric profiles of the cluster mass
components; see Table 5 in G15 and Table 4 in G16), according to the value of the minimum $\chi^{2}$. As
shown in Figure 17 by G15 and Figure 6 by G16, it
is remarkable that all tested mass parametrizations lead to
statistical and systematic relative errors of only a few percent for
the cluster total surface mass density and cumulative projected
mass. The latter has also been found to be in very good agreement with the
estimates from independent mass diagnostics, e.g. those from weak
lensing, dynamical and and X-ray observations (see e.g.,
G15; \citealt{bal16}; \citealt{cam17}). Moreover, in the
modeling of a galaxy cluster, the inclusion of the different mass
components (i.e., extended dark-matter halos, cluster members, and
possibly hot gas; see e.g., \citealt{bona17}; \citealt{ann17})
provides a good approximation of the first-order lensing effects
from the mass distributions in the regions adjacent to where the time
delays are measured (i.e., the possible effect of the environment). 
In summary, if extensive multi-color and spectroscopic information is
available in lens galaxy clusters, like in MACS 1149, it is possible
to construct robust mass maps (see G15; \citealt{cam17b}; \citealt{lag17}). We demonstrate the feasibility of using SN Refsdal for measuring $H_0$ with high statistical precision; the full systematic analysis will be in future work when the final time-delay measurements from the light curve monitoring becomes available. 
We remark that our first tests adding to the model a uniform sheet of mass at the cluster redshift (free to vary, with a flat prior) and optimized together with all the other model parameters result in median values that are very close to 0 and in $H_{0}$ probability distribution functions that are just slightly broader than those (presented above) without this extra mass component. Based on our previous studies (see e.g., \citealt{chi18} on the influence of line-of-sight structures on lensing modeling) and additional preliminary results, we anticipate that the systematic effects in MACS 1149 could be controlled to a level similar to the statistical uncertainties given the exquisite data set in hand, making time-delay cluster lenses a potentially competitive cosmological probe.

Finally, we comment briefly on the recent work by \citet{veg17}, where
an estimate of the value of $H_{0}$ has been obtained by combining the
time delay predictions of the different groups who participated in the
blind analysis on the reappearence of SN Refsdal published by
\citet{tre15b}. We notice that not all models perform equally well in
reproducing and predicting the positions, fluxes and time delays of
the multiple images of SN Refsdal (see Figures 7 and 8 in
\citealt{rod16} and Figures 2 and 3 in \citealt{kel16}), so it is not
very meaningful to assign the same weight to all model
predictions. In fact, some of the models cannot reconstruct the expected topology of the arrival time-delay surface near the multiple images of SN Refsdal (see Figure 8 in \citealt{tre15b}), and they do not produce images at those positions.
Furthermore, we remark that in a strong lensing model, with a set of multiply-imaged sources at different redshifts, the values of the cosmological parameters and those defining the total mass distribution of the lens are not independent and they cannot be considered separately in deriving predictions (e.g., time delays and flux ratios of multiple images). Contrary to what we have done in the analysis presented here, \citet{veg17} simply rescale the model-predicted quantities varying only the value of $H_{0}$ and keeping the total mass models of the cluster fixed. Therefore, the results obtained with this methodology are likely to underestimate the uncertainty on the value of $H_{0}$, and possibly introduce biases, since they neglect the covariance between $H_{0}$ and the cluster model parameters. The work by \citet{zit14} confirms the presence of a bias in the values of the cosmological parameters when they are inferred by applying a fixed lens mass model for correcting the luminosity distances of lens-magnified Type Ia SNe.


\section{Conclusions}
\label{sec:conc}

We have shown that it is possible to measure precisely the value of the Hubble constant by using a large set of observed images from spectroscopic multiply-lensed sources and the measured time delays between the multiple images of a variable source in a lens galaxy cluster. We have modeled the extraordinary photometric and spectroscopic data in the HFF galaxy cluster MACS J1149.5$+$2223 and shown that the value of $H_{0}$ can be inferred, without intermediate calibrations and any priors on the values of $\Omega_{m}$ and $\Omega_{\Lambda}$, with a 6\% percent statistical error in flat $\Lambda$CDM models, if the time delay of the latest image of SN Refsdal (SX) is known with a 3\% uncertainty. The precision on the $H_{0}$ value should be even higher, once the final time delay of SX, with the expected $\approx$1-2\% relative precision, becomes available. At that point, our best estimate of $H_{0}$, based on the model of SN Refsdal detailed here, will be presented. We have tested this method, originally proposed by \citet{ref64}, in more general cosmological models and have found that it can also provide a new way to measure the values of $\Omega_{m}$ and $\Omega_{\Lambda}$ that is competitive with other standard techniques. When applied to other strong-lensing systems, with high quality data, that are already available or that are expected to be discovered in forthcoming deep and wide surveys, this will become an important and complementary tool to measure the expansion rate and the geometry of the Universe.

\smallskip

\acknowledgments
C.G. acknowledges support by VILLUM FONDEN Young Investigator Programme through grant no.~10123. 
S.H.S.~thanks the Max Planck Society for support through the Max Planck Research Group.
T.T. acknowledges support by the Packard Foundation through a Packard Research Fellowship and by NASA through grant HST-GO-14199.
This work is based in large part on data collected in service mode at ESO VLT, through the Director's Discretionary Time Programme 294.A-5032.
The CLASH Multi-Cycle Treasury Program is based on observations made with the NASA/ESA {\it Hubble Space Telescope}. The Space Telescope Science Institute is operated by the Association of Universities for Research in Astronomy, Inc., under NASA contract NAS 5-26555. ACS was developed under NASA Contract NAS 5-32864.

\clearpage

\clearpage

\end{document}